\documentclass[a4paper]{article}

\newcommand{\s}   {\mathrm{s   }}
\newcommand{\cm}  {\mathrm{cm  }}
\newcommand{\ns}  {\mathrm{ns  }}
\newcommand{\mev} {\mathrm{MeV }}
\newcommand{\gev} {\mathrm{GeV }}
\newcommand{\tev} {\mathrm{TeV }}
\newcommand{\fb}  {\mathrm{fb  }}
\newcommand{\intL}{\int \mathcal{L} \mathrm{d}t}
\newcommand{\HToZZ}{H\to ZZ^{\star}\to\ell\ell\ell\ell}
\newcommand{\HToGG}{H\to\gamma\gamma}
\newcommand{\ggToLLLL}{gg\to\ell\ell\ell\ell}

\usepackage{graphicx}
\usepackage[affil-it]{authblk}

\usepackage[utf8]{inputenc}
\usepackage[T1]{fontenc}
\usepackage{lmodern} 

\begin{document}

\title{Study of systematic errors on the scalar boson mass}
\author{A. Randle-Conde, on behalf of the ATLAS and CMS collaborations}
\affil{IIHE-ULB, Université Libre de Bruxelles, Belgium}
\date{Talk presented at the International Workshop on Future Linear Colliders (LCWS14), Belgrade, Serbia, 6-10 October 2014.}

\maketitle


\section{Introduction}

In this talk I presented studies of the scalar boson mass resolution at the LHC.  I described in detail the limiting factors on resolution, the methods used to determine the mass resolution, and the statistical sample size required for these results for the $\HToZZ$ and $\HToGG$ final states.  I finished the talk by comparing projections for various linear collider scenarios.

\section{Physics at the LHC}

The LHC is located at CERN, Geneva, Switzerland.  The LHC is a $27~\mathrm{km}$ circular accelerator consisting of two counter rotating beams, capable of accelerating protons and lead ions up to energies of $7~\tev$ per proton and $2.76~\tev$ per nucleon.  The proton-proton beam conditions at the LHC are shown in table \ref{tab:LHC_beam}.
Due to the nature of the beams at the LHC the detector environment is different when compared to one at a linear collider.

\begin{table*}[!bh]
  \begin{center}
    \begin{tabular}{ccccc}
      \hline
      Parameters                                                   & Design & $2011$ & $2012$    & $2015$ \\
      \hline
      Beam energy ($\tev$)                                         & $7$    & $3.5$  & $4$       & $6.5$  \\
      Particles per bunch ($10^{11}$)                              & $1.15$ & $1.5$  & $1.6-1.7$ & $1.15$ \\
      Number of bunches                                            & $2808$ & $1380$ & $1374$    & $2590$ \\
      Bunch spacing ($\ns$)                                        & $25$   & $50$   & $50$      & $25$   \\
      Interactions per crossing ($\langle\mu\rangle$) & $~20$  & $6-11$ & $~40{}^\dagger$     & $~50$  \\
      Peak luminosity ($10^{34}~\cm^{-2}\s^{-1}$)                  & $1.0$  & $0.36$ & $0.77$    & $1.7$  \\
      \hline
    \end{tabular}
    \caption{The parameters for the LHC, for different conditions: design conditions, the $2011$ and $2012$ data taking, and expected conditions for $2015$ data taking \cite{LHCDesignReport} \cite{LHCStatus}.  (${}^\dagger$ Maximum mean value)}
    \label{tab:LHC_beam}
  \end{center}
\end{table*}

\begin{figure*}[hbtp]
  \begin{center}
    \begin{tabular}{cc}
      \includegraphics[width=0.4\textwidth]{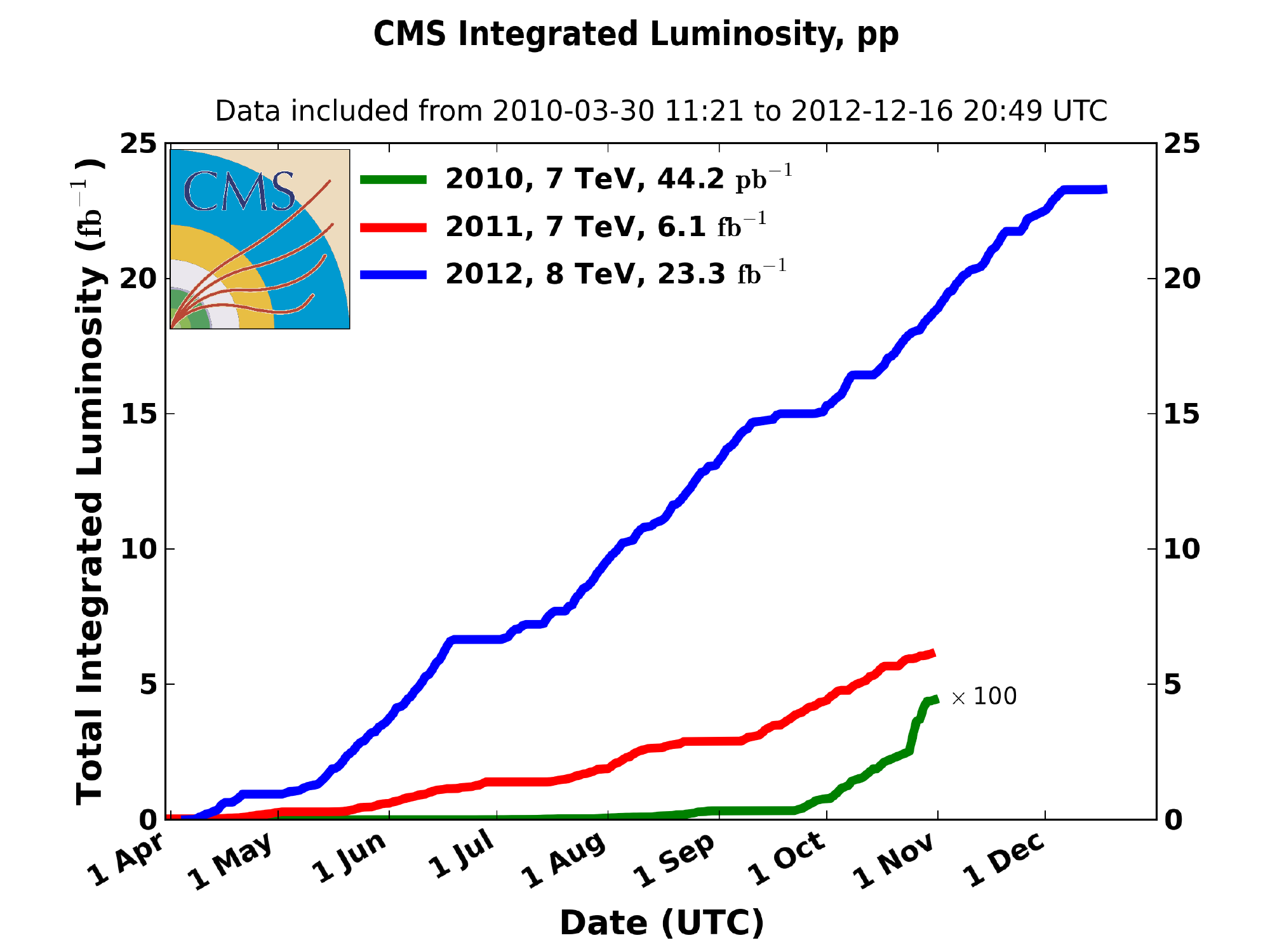} &
      \includegraphics[width=0.4\textwidth]{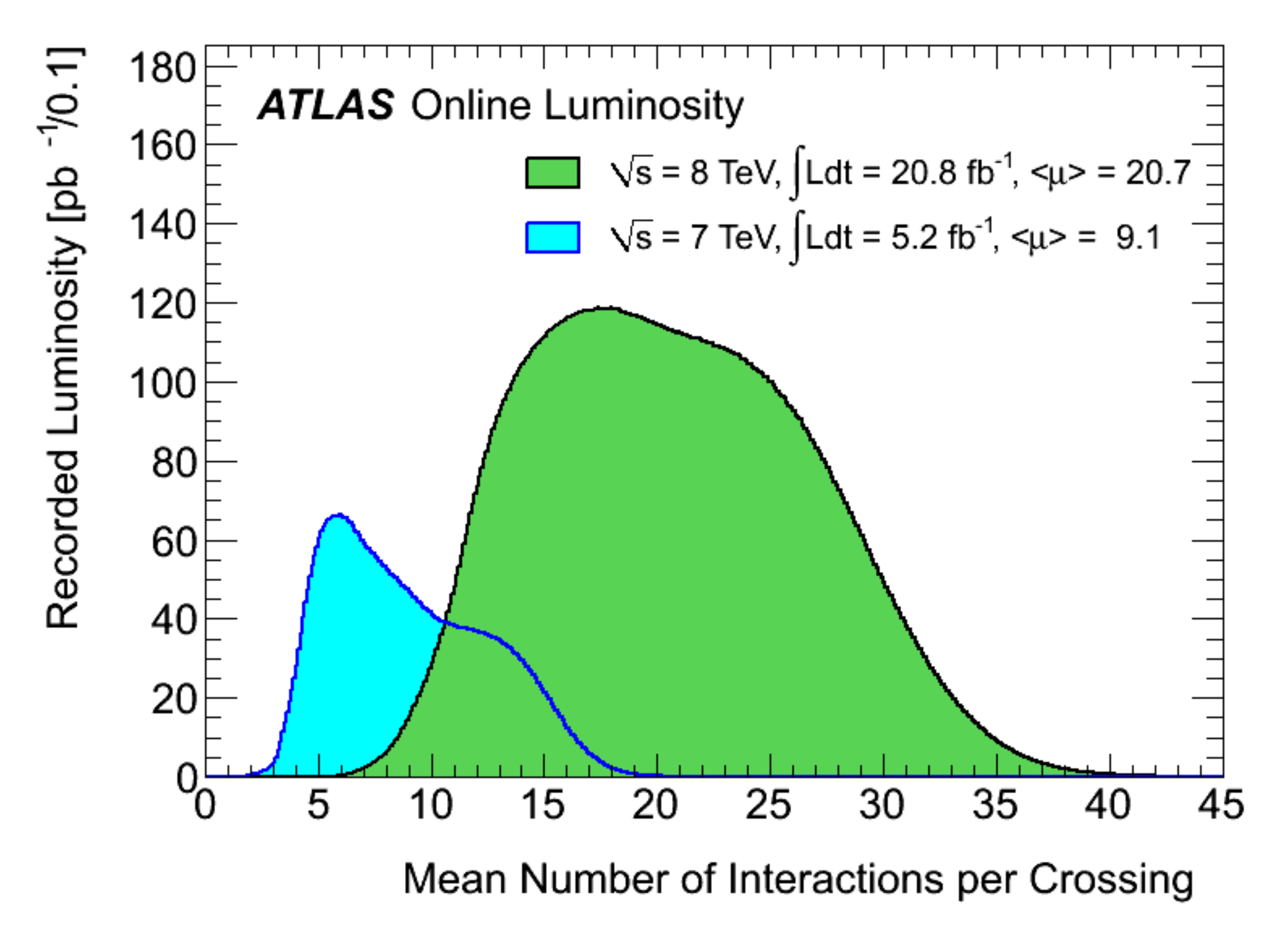}
    \end{tabular}
    \caption{The recorded luminosity (for CMS) \cite{CMSLumi} (left) and recorded pileup (for ATLAS) \cite{ATLASPileup} (right) showing typical values delivered by the LHC.}
    \label{fig:LHC_lumi}
  \end{center}
\end{figure*}

The ATLAS \cite{ATLASDetector} and CMS \cite{CMSDetector} detectors are two general purpose detectors located on the LHC ring.  Each consists of several subdetectors, including an inner tracking system which is used to reconstruct tracks left by charged particles.  Each detector also contains an electromagnetic calorimeter which is used to reconstruct electrons and photons, and an hadronic calorimeter which is used to reconstruct hadronic jets.  The outer sections of the detectors consist of various muon systems, which are used to reconstruct muon trajectories.  The CMS detector has a superconducting solenoid magnet which surrounds the inner tracking system and calorimeters, with a return yoke in the outer detector.  The ATLAS detector has a superconducting solenoid magnet which surrounds the inner tracking system, and toroidal magnets in the outer detector.

At the LHC, the scalar boson can be mainly produced via four production modes: gluon-gluon fusion, vector boson fusion, associated production, and $t\bar{t}$ associated production.  The dominant decay modes are $H\to b\bar{b}$, $H\to WW$, $H\to \tau\tau$, however these all have poor mass resolution.  The rare decay modes $H\to ZZ^{\star}$ and $\HToGG$ give fine mass resolution, whereas the modes $H\to Z\gamma$ and $H\to\mu\mu$ are not yet accessible at the LHC due to large backgrounds and small branching fractions.  The scalar boson linewidth is not directly accessible at the LHC.

\section{$\HToZZ$ final state}

The decay $\HToZZ$, known as the ``golden mode'', has very fine mass resolution.  The mass measurements for ATLAS and CMS are presented in table \ref{tab:combination}.  The systematic uncertainties are dominated by lepton energy and momentum scale uncertainties, so dedicated studies are performed using narrow resonances to estimate these uncertainties.

\begin{figure*}[hbtp]
  \begin{center}
    \begin{tabular}{cc}
      \includegraphics[width=0.5\textwidth]{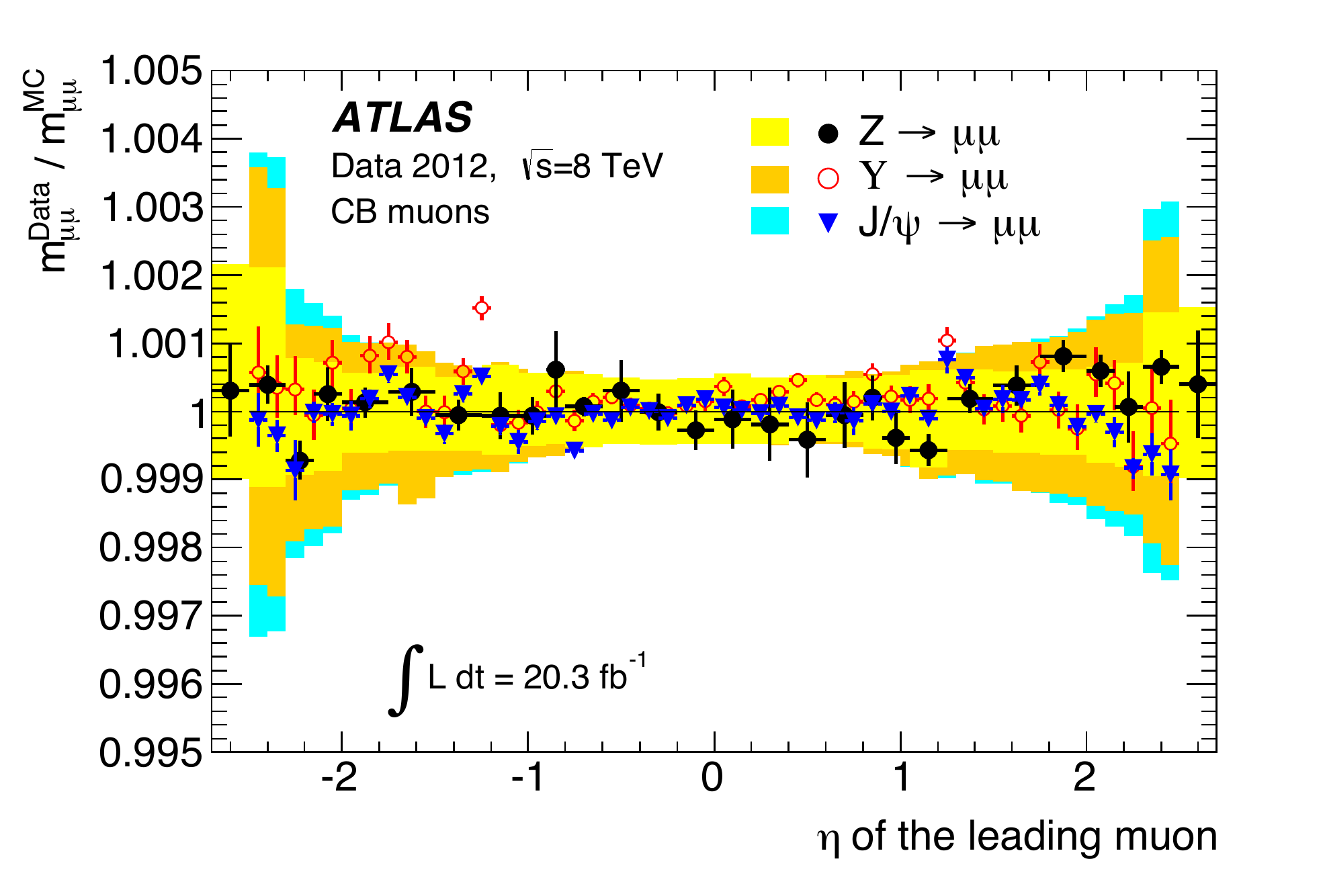} &
      \includegraphics[width=0.5\textwidth]{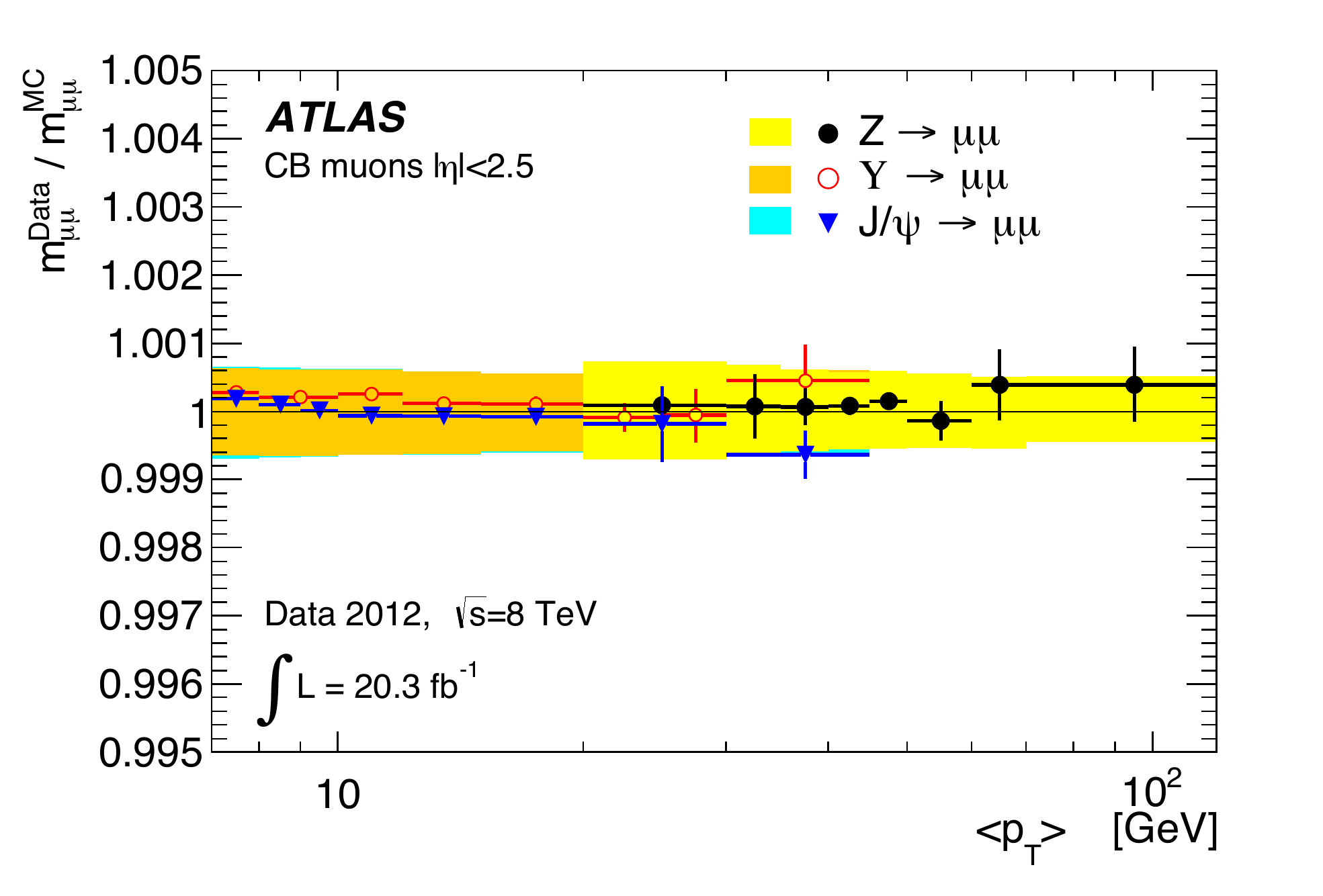}
    \end{tabular}
    \caption{The ratios of reconstructed dimuon invariant masses in data to the corrected masses in simulation for various resonances for the ATLAS muon calibration as a function of $\eta$ (left) and $p_T$ (right) \cite{ATLASCombo}.}
    \label{fig:muons_ATLAS}
  \end{center}
\end{figure*}

The ATLAS collaboration performs a study using the $Z$ and $J/\psi$ resonances for muon calibration.  The Monte Carlo simulation (MC) is smeared and scaled to match the data, binning in pseudorapidity ($\eta$) and transverse momentum ($p_T$) of the muons.  Muons are split into three categories according to which detector subsystems are used in reconstructions.  The momentum scale uncertainties are $0.02\% (|\eta|=0)- 0.2\% (|\eta|>2)$ ($<0.1\%$ overall) for inner detector muons, $<0.2\%$ for outer detector muons, and $0.04\%$ (barrel) $-0.2\% (|\eta|>2)$ for combined muons.  The ratios of reconstructed dimuon invariant masses in data to the corrected masses in simulation for various resonances for the ATLAS muon calibration are shown in figure \ref{fig:muons_ATLAS}.

\begin{figure*}[hbtp]
  \begin{center}
    \begin{tabular}{cc}
      \includegraphics[width=0.4\textwidth]{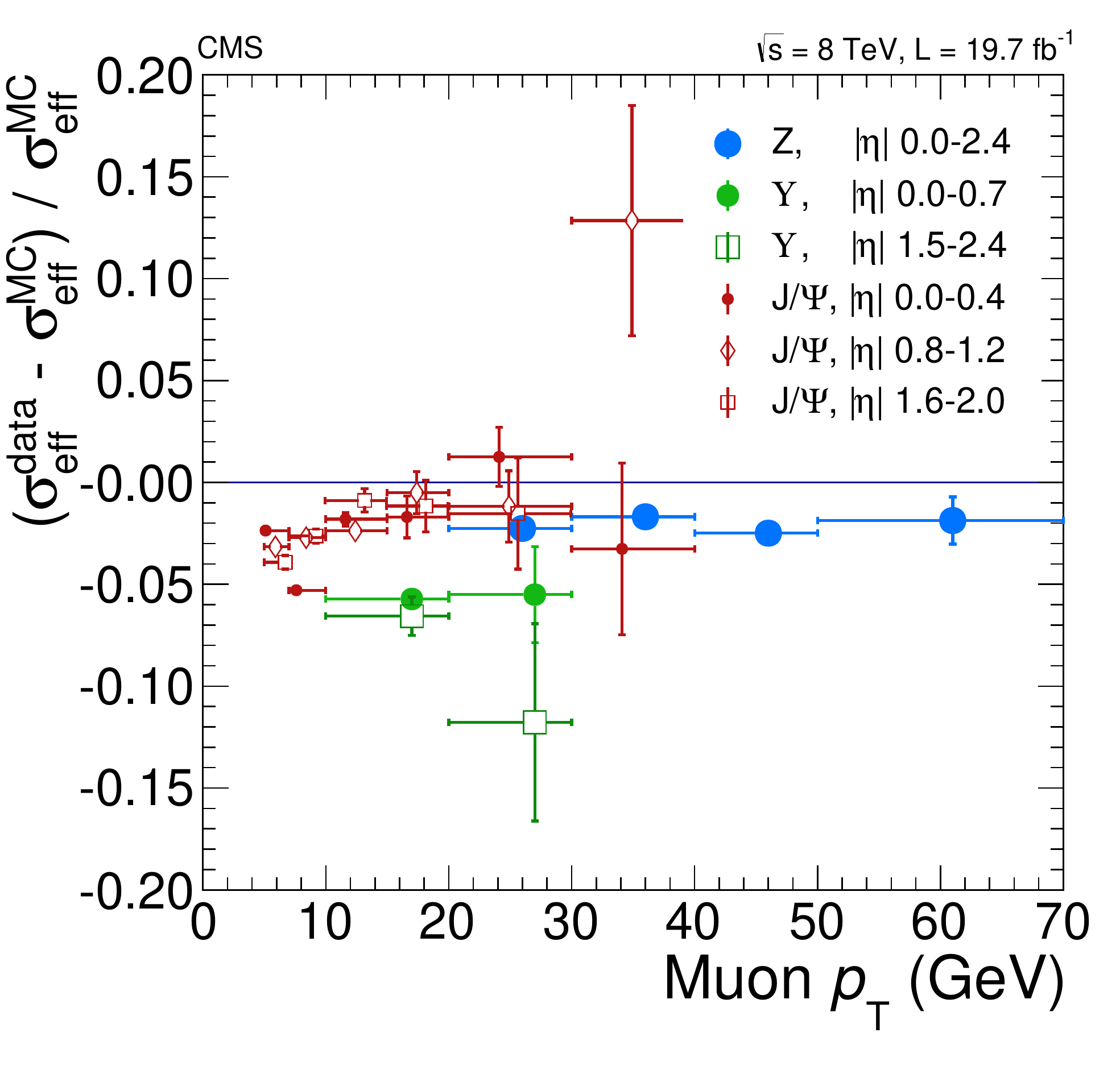} &
      \includegraphics[width=0.4\textwidth]{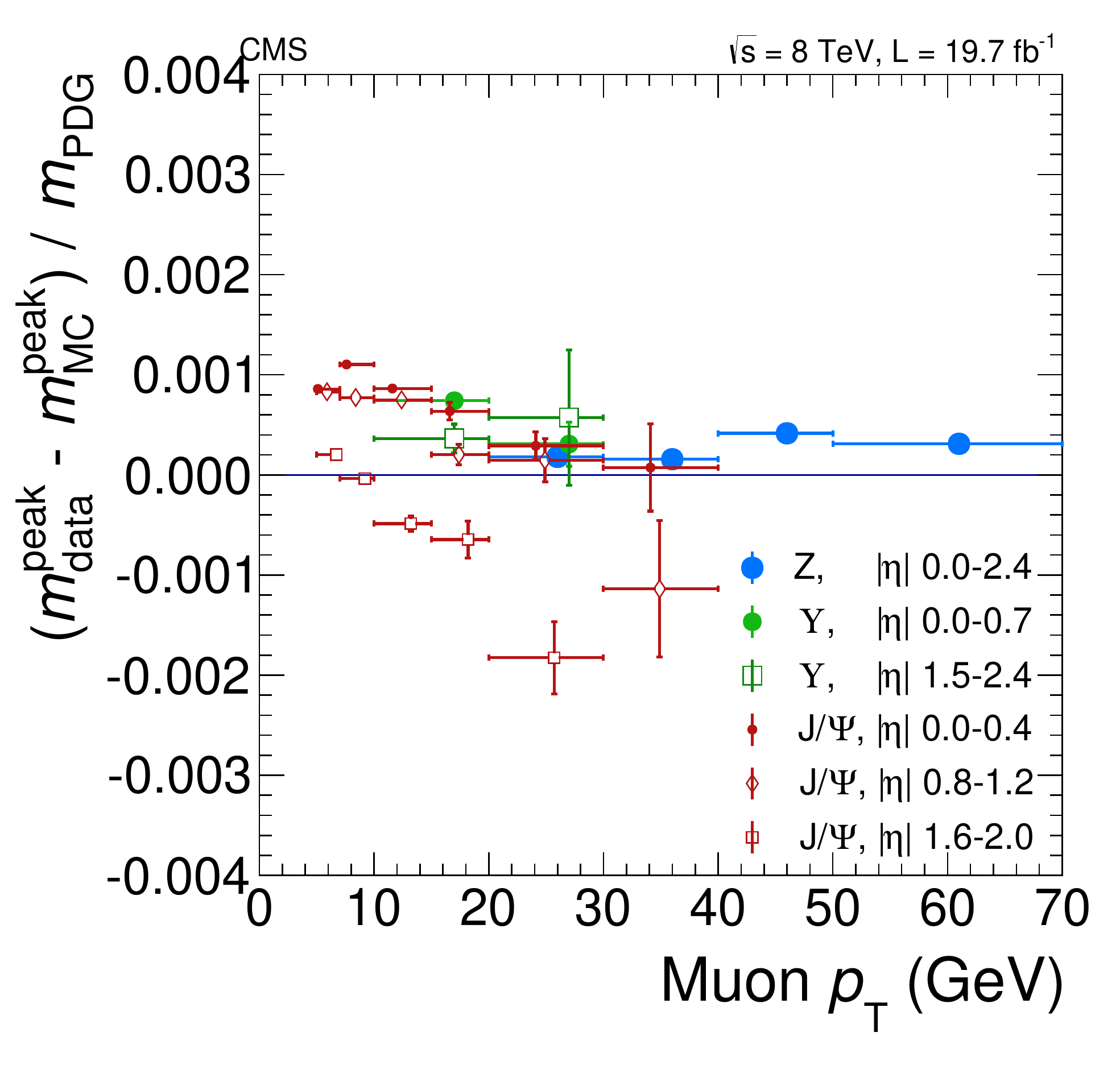}
    \end{tabular}
    \caption{The mass resolution (left) and scale factor (right) for the CMS muon calibration as a function of $p_T$ \cite{CMSHToZZ}.}
    \label{fig:muons_CMS}
  \end{center}
\end{figure*}

The CMS collaboration performs a study using the $Z$, $J/\psi$, and $\Upsilon$ resonances for muon calibration.  MC is smeared and scaled to match the data at the $Z$ resonance using a $Z$ lineshape convoluted with a Gaussian function.  Momentum scale and resolution measurements are made using biases which are determined using shifts in the position of the $Z$ peak.  Events are fitted to the $Z$, $J/\psi$, and $\Upsilon$ resonances, averaging over $\eta$ and $p_T$.  The uncertainty in the peak position is $0.1\%$ and the uncertainty in the peak width is $5\%$.  The resolution uncertainties for muons, and shift in peak position are shown in figure \ref{fig:muons_CMS}.

For both experiments the typical momentum scale uncertainty is of the order of $~0.1\%$ or better per muon.  The momentum scale uncertainty measurements use all of the LHC Run I datasets, and are statistically limited.  The uncertainties and sample sizes are shown in table \ref{tab:muon_momentumScale}.

\begin{table*}
  \begin{center}
    \begin{tabular}{ccccc}
      \hline
                                           & ATLAS               & CMS                           \\
      \hline
      Momentum scale uncertainty           & $0.04\%-0.2\%$      &       -                       \\
      Mass uncertainty                     &          -          & $0.1\%$ in $4\mu$ final state \\
      \hline
      Number of $Z\to\mu\mu$ events        & $ \sim 9,000,000^1$ & $\sim 14,000,000^{1,2}$       \\
      Number of $J/\psi\to\mu\mu$ events   & $\sim 17,000,000^1$ & $\sim 27,000,000^2$           \\
      Number of $\Upsilon\to\mu\mu$ events & $ \sim 5,000,000^2$ & $\sim 15,000,000^2$           \\
      \hline
    \end{tabular}
    \caption{The contribution of momentum scale uncertainties to mass uncertainties for ATLAS \cite{ATLASCombo} and CMS \cite{CMSHToZZ} from the $\HToZZ$.  Samples labelled ${}^1$ are used to correct MC samples.  Samples marked ${}^2$ are used to validate MC samples.}
    \label{tab:muon_momentumScale}
  \end{center}
\end{table*}

\begin{figure*}[hbtp]
  \begin{center}
    \begin{tabular}{ccc}
      \includegraphics[width=0.3\textwidth]{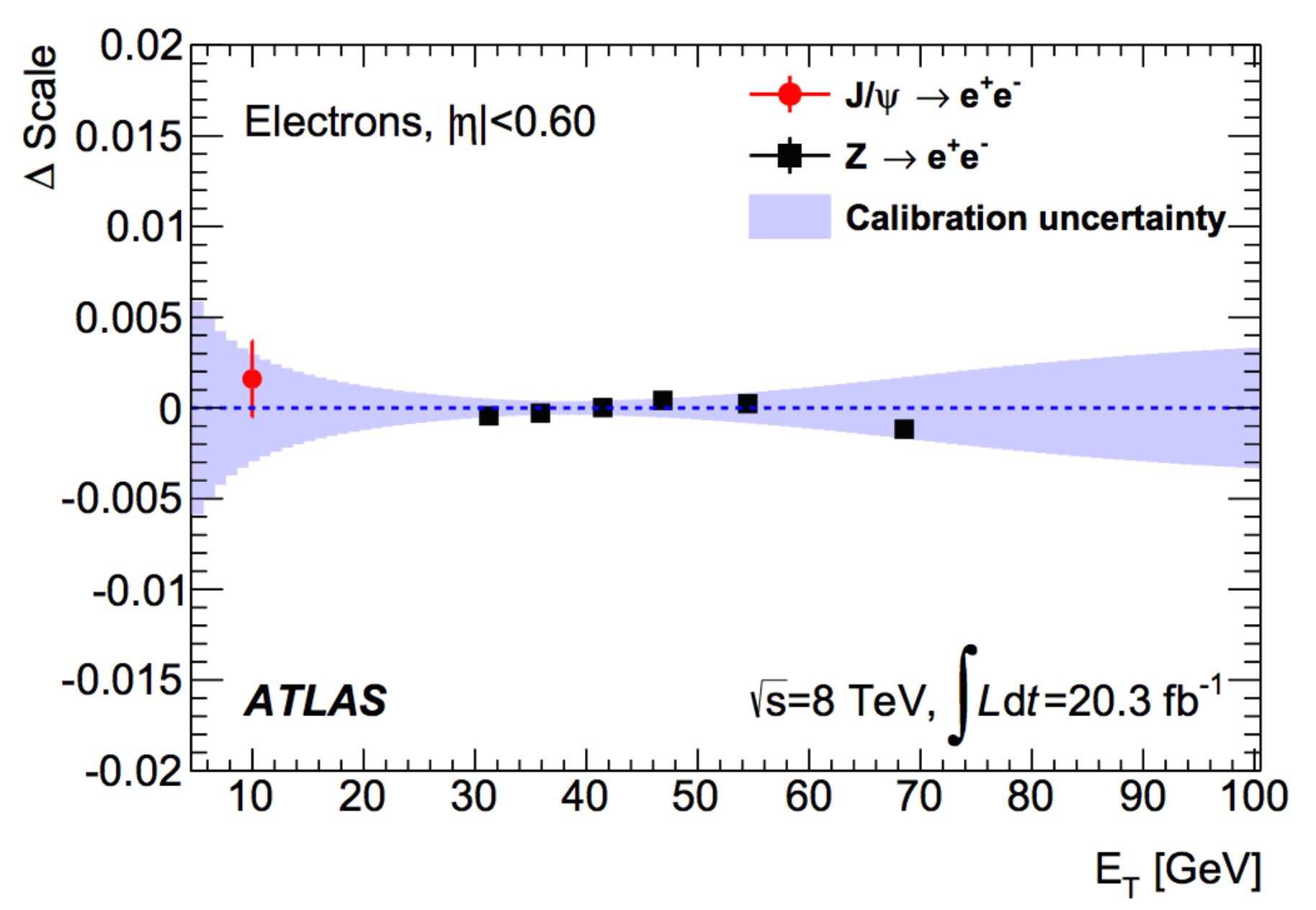} &
      \includegraphics[width=0.3\textwidth]{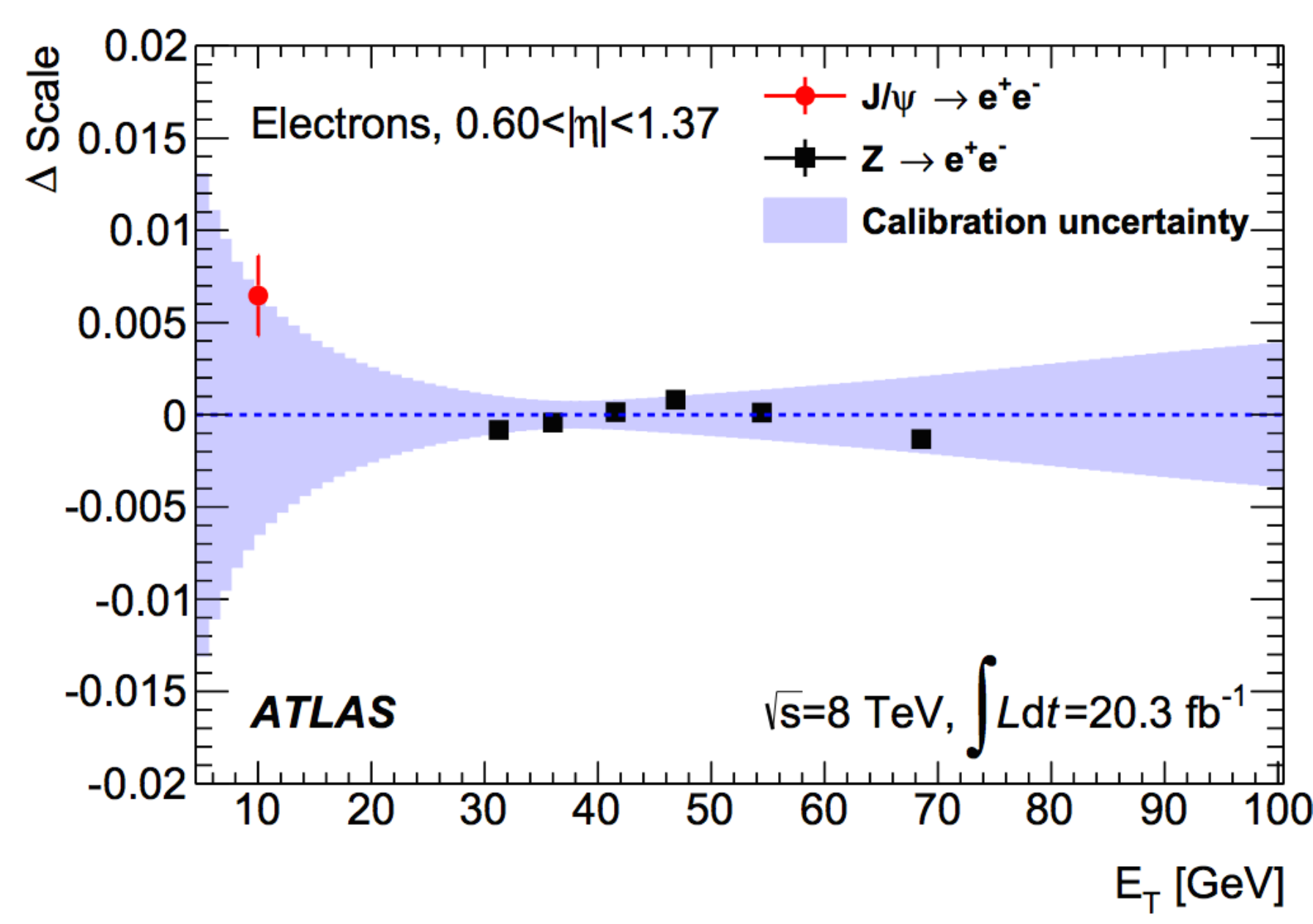} &
      \includegraphics[width=0.3\textwidth]{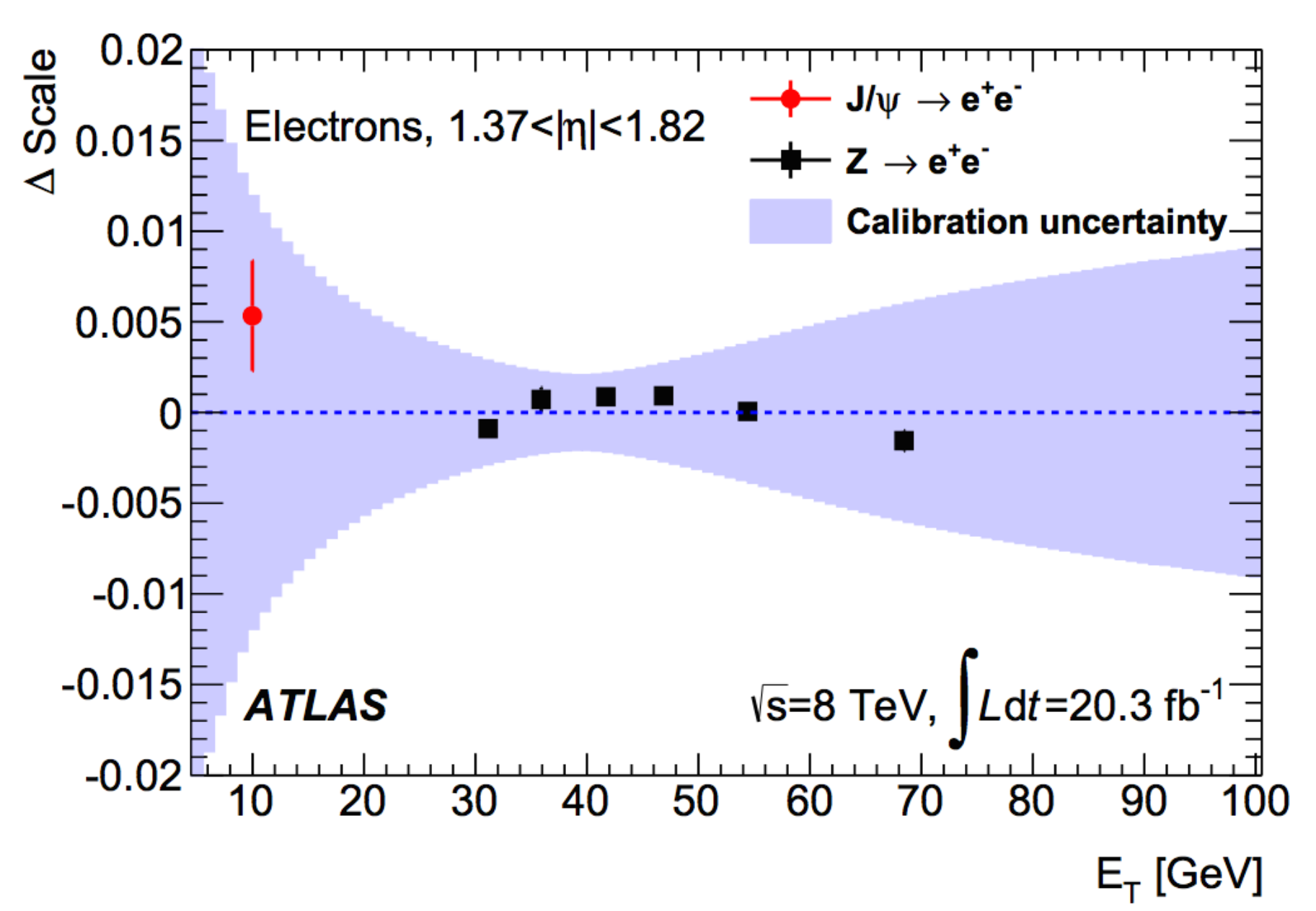}
    \end{tabular}
    \caption{The relative scale differences for electrons, measured and nominal, for ATLAS as a function of $E_T$ for different $|\eta|$ ranges \cite{ATLASCombo}.}
    \label{fig:electrons_ATLAS}
  \end{center}
\end{figure*}

The ATLAS collaboration performs a study using the $Z$ and $J/\psi$ resonances for electron calibration.  Electron candidates are first calibrated using a multivariate discriminant using MC samples.  The material budget of the detector is taken into account by measuring the energy depositions of the first and second layers of the electromagnetic calorimeter.  Uniformity corrections are applied to take variations with respect to azimuthal angle, time, and pileup into account.  Longitudinal scale variations are taken into account in data, and then data and MC are compared to obtain MC based calibrations.  The $Z\to ee$ peak is used to calibrate the energy scale in data, and the transverse energy, $E_T$, in MC.  These calibrated electrons are validated using $J/\psi\to ee$ and $Z\to ee\gamma$ samples.  The relative scale differences for electrons for different $\eta$ ranges are shown in figure \ref{fig:electrons_ATLAS}.  The electron energy scale uncertainties at two $E_T$ points are shown in table \ref{tab:electrons_ATLAS}.

\begin{table*}
  \begin{center}
    \begin{tabular}{ccccc}
      \hline
      $E_T$ ($\gev$) & $|\eta|<1.37$ & $1.37<|\eta|<1.82$ & $1.82<|\eta|$ \\
      \hline
      $\sim 40\gev$  & $0.04\%$      & $0.2\%$            & $0.05\%$      \\
      $\sim 11\gev$  & $0.4-1\%$     & $1.1\%$            & $0.4\%$       \\
      \hline
    \end{tabular}
    \caption{The energy scale uncertainties for electrons at two transverse energy points for ATLAS \cite{ATLASElectronsPhotons}.}
    \label{tab:electrons_ATLAS}
  \end{center}
\end{table*}

\begin{figure*}[hbtp]
  \begin{center}
    \begin{tabular}{cc}
      \includegraphics[width=0.4\textwidth]{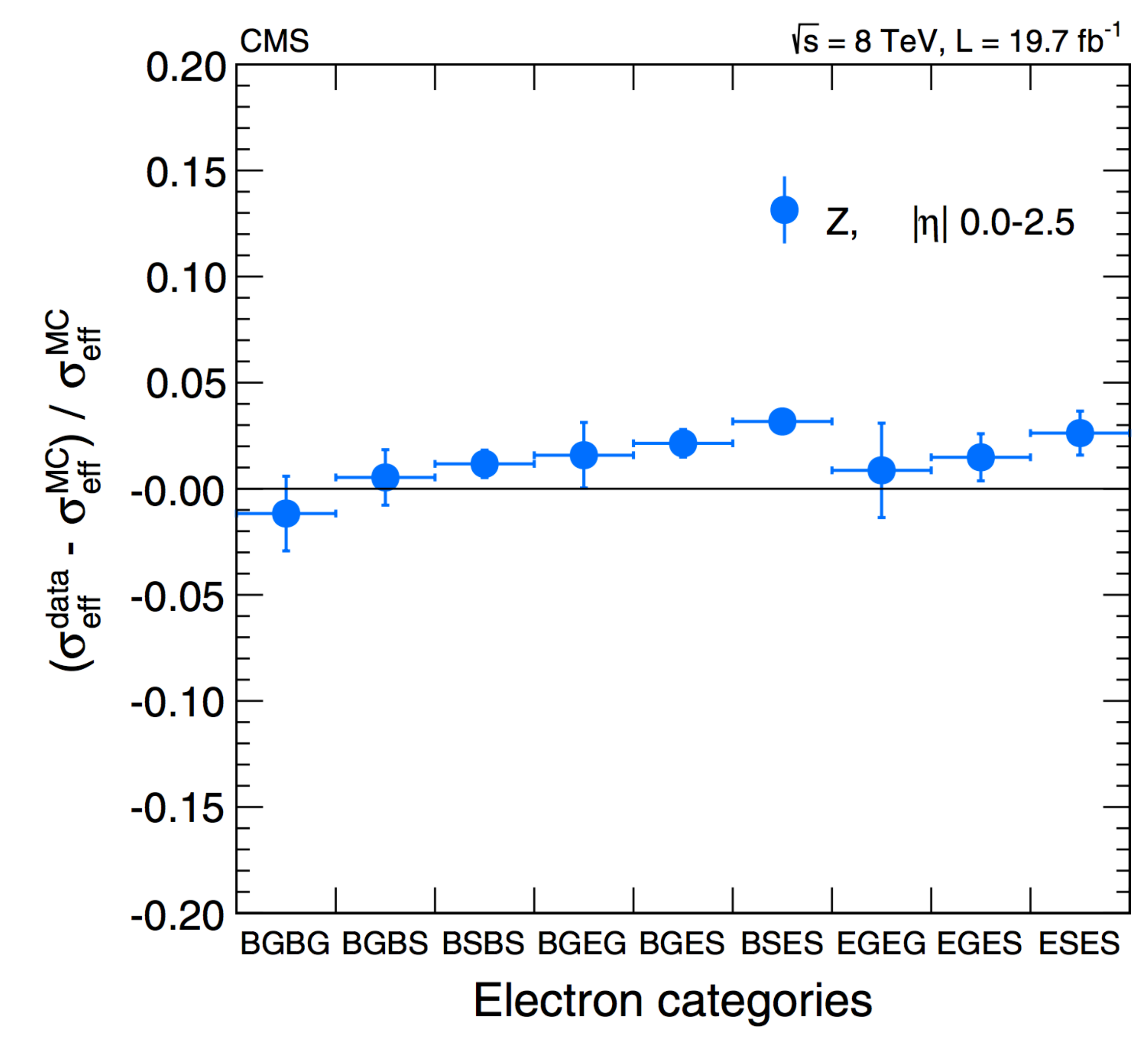} &
      \includegraphics[width=0.4\textwidth]{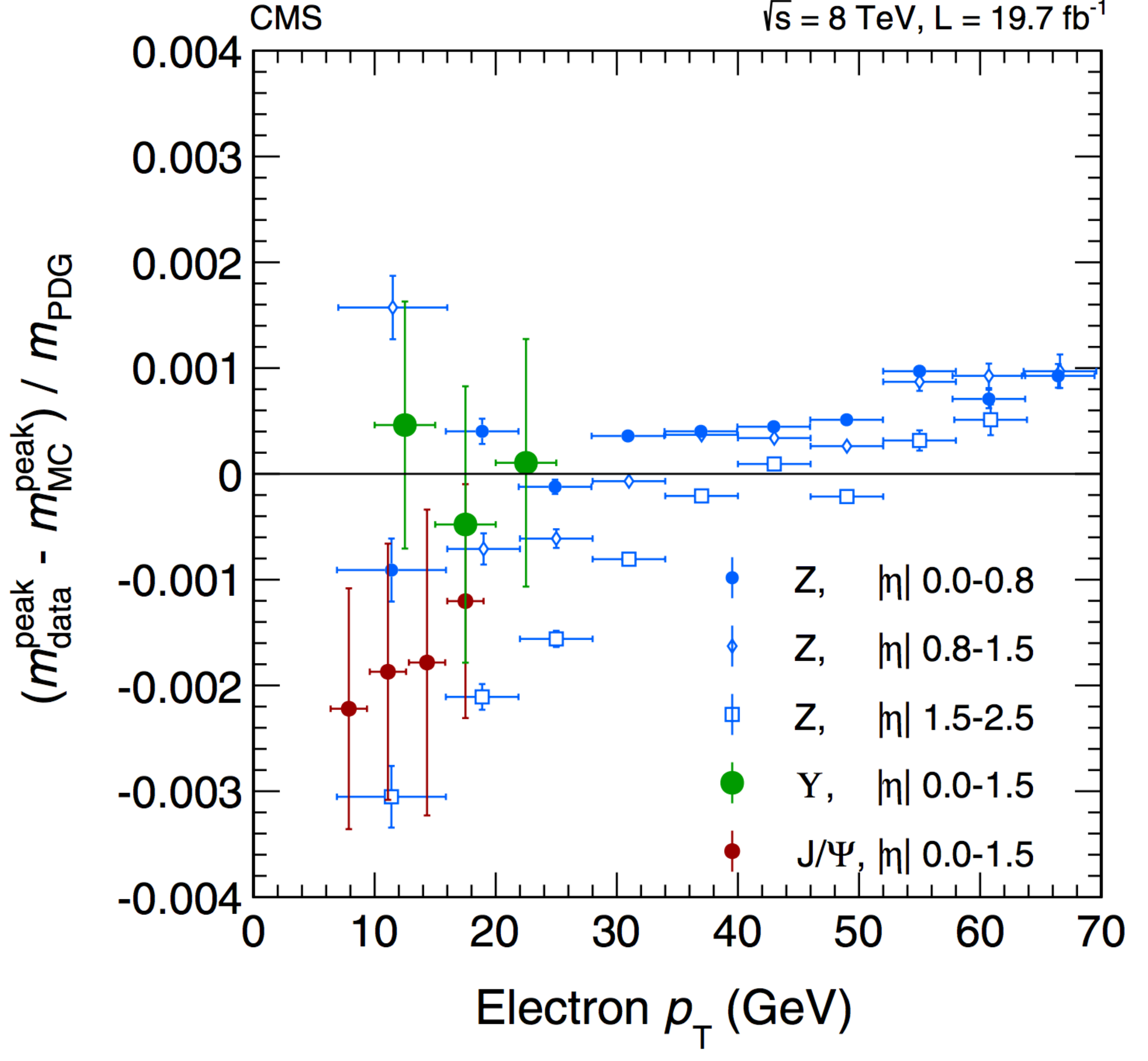}
    \end{tabular}
    \caption{The efficiency uncertainties for different electron categories (left) and the shift in mass peak position (right) between data and MC.  The electron categories are: barrel (B), endcap (E), golden (G), showering (S) \cite{CMSHToZZ}.}
    \label{fig:electrons_CMS}
  \end{center}
\end{figure*}

The CMS collaboration performs a study using the $Z$, $J/\psi$, and $\Upsilon$ resonances for electron calibration.  Electron momentum scale corrections are determined using the difference in the $Z$ peak position between data and MC.  Time dependence is implicit in the corrections, to take the transparency losses of the crystals into account.  The transverse momentum dependence is taken into account using linearity corrections, using $J/\psi$ and $\Upsilon$ resonances to validate the $p_T<20 ~\gev$ region.  Energies of single electrons are then smeared with a Gaussian function with a width of $\Delta\sigma$, where $\Delta\sigma$ is the difference in resolution between data and MC.  Events are categorised according to the $p_T$ and $\eta$, and fitted to mass peaks of the $Z$, $J/\psi$, and $\Upsilon$ resonances.  The systematic uncertainties of the peak position are $0.1\%$ for the $2e2\mu$ final state, and $0.3\%$ for the $4e$ final state, and the systematic uncertainty of the peak width is $1.2-4\%$.  The efficiency uncertainties for different electron categories, and shift in peak position are shown in figure \ref{fig:electrons_CMS}.

For both experiments the typical momentum scale uncertainty is approximately $0.5\%$ or better per electron.  The momentum scale uncertainty measurements use all of the LHC Run I datasets, and are statistically limited.  The uncertainties and sample sizes are shown in table \ref{tab:electron_momentumScale}.

\begin{table*}
  \begin{center}
    \begin{tabular}{ccccc}
      \hline
                                        & ATLAS                            & CMS                             \\
      \hline
      Momentum scale uncertainty        & $E_T\sim 40\gev :0.03\%-0.05\%$  &       -                         \\
                                        & $E_T\sim 10\gev :0.4\%-1.1\%$    &                                 \\
      Mass uncertainty                  &          -                       & $0.3\%$ in $4e$ final state     \\
                                        &          -                       & $0.1\%$ in $2e2\mu$ final state \\
      \hline
      Number of $Z\to ee$ events        & $\sim 6,600,000^1$ & $\sim 10,000,000^{1,2}$ \\
      Number of $J/\psi\to ee$ events   & $  \sim 300,000^1$ &           $\sim 5000^2$ \\
      Number of $Z\to ee\gamma$ events  & $  \sim 200,000^2$ &                       - \\
      Number of $\Upsilon\to ee$ events &                  - &         $\sim 25,000^2$ \\
      \hline
    \end{tabular}
    \caption{The contribution of momentum scale uncertainties to mass uncertainties for ATLAS \cite{ATLASCombo} and CMS \cite{CMSHToZZ} from the $\HToZZ$.  Samples labelled ${}^1$ are used to correct MC samples.  Samples marked ${}^2$ are used to validate MC samples.}
    \label{tab:electron_momentumScale}
  \end{center}
\end{table*}


Due to the low branching fraction there are very few signal events in the $\HToZZ$ final state.  As a result the mass variations between events is significant and the mass resolution can be improved by taking per-event mass uncertainties into account.  Both ATLAS and CMS use per event mass uncertainties, using likelihoods of the form $P(\mathcal{D}_m|m_{H})$, where $\mathcal{D}_m$ is the per event mass uncertainty estimation.  Nuisance parameters are fixed to their best-fit values, and the likelihood analysis is performed a second time.


The mass resolution is much larger than the predicted width of the scalar boson ($\Gamma_H=4.15 ~\mev$ as $m_H=125.5 ~\gev$), so indirect measurements of the width have been developed.  Comparing on-shell and off-shell contributions to the $\ggToLLLL$ processes allows ATLAS and CMS to obtain indirect limits of $\Gamma_H<24 ~\mev$ \cite{ATLASWidth} and $\Gamma_H<22 ~\mev$ \cite{CMSWidth} respectively, approximately an order of magnitude lower than the mass resolution.  The dominant uncertainties on the width estimation are related to the interference of $gg\to H\to ZZ^{\star}$ and $gg\to ZZ^{\star}$, and are of the order of $20\%$.

\section{$\HToGG$ final state}

The decay $\HToGG$ has large statistics for a $H$ boson mass in the region $m_H<140 ~\gev$, however the standard model backgrounds are very large in this region.  For both ATLAS and CMS the detector response can be calibrated using $Z\to ee$ (and $Z\to ee\gamma$ for ATLAS) events.  Further sensitivity is gained by categorising the events based on the properties of the photons.  The mass measurements for ATLAS and CMS are shown in table \ref{tab:combination}.


The response of the detectors to photons are very similar to those of electrons.  As a result the methods used to calibrate photons are very similar to those used to calibrate electrons.  The mass uncertainty in the measurement for ATLAS is dominated by photon energy scale uncertainties.  The uncertainties are summarised in table \ref{tab:photons_ATLAS}.

\begin{table*}
  \begin{center}
    \begin{tabular}{ccc}
      \hline
      Source of                      & Unconverted  & Converted     \\
      uncertainty                    & photons      & photons       \\
      \hline
      $Z\to ee$ calibration         & $0.02-0.11\%$ & $0.02-0.11\%$ \\
      Calorimeter cell non-linearty & $0.09-0.39\%$ & $0.06-0.29\%$ \\
      Layer calibration             & $0.11-0.16\%$ & $0.05-0.10\%$ \\
      Identification material       & $0.06-0.10\%$ & $0.05-0.06\%$ \\
      Other material                & $0.07-0.35\%$ & $0.04-0.20\%$ \\
      Conversion reconstruction     & $0.02-0.05\%$ & $0.02-0.06\%$ \\
      Lateral shower shape          & $0.04-0.07\%$ & $0.09-0.19\%$ \\
      \hline
      Total                         & $0.23-0.59\%$ & $0.21-0.47\%$ \\
      \hline
    \end{tabular}
    \caption{Sources of systematic uncertainty for photon reconstruction at ATLAS \cite{ATLASCombo}.}
    \label{tab:photons_ATLAS}
  \end{center}
\end{table*}

For CMS the mass uncertainties in the $\HToGG$ final state are dominated by differences between electron and photon reconstruction, linearity of the energy scales, and energy scale calibration and resolution.  These, and other uncertainties, are shown in table \ref{tab:photons_CMS}.

\begin{table*}
  \begin{center}
    \begin{tabular}{ccc}
      \hline
      Source of uncertainty                   & Uncertainty ($\gev$) \\
      \hline
      Electron-photon differences             & $0.10$      \\
      Linearity of the energy scale           & $0.10$      \\
      Energy scale calibration and resolution & $0.05$      \\
      Other contributions                     & $0.04$      \\
      \hline
      Total                                   & $0.15$      \\
      \hline
    \end{tabular}
    \caption{Sources of systematic uncertainty for photon reconstruction at CMS \cite{CMSHToGG}.}
    \label{tab:photons_CMS}
  \end{center}
\end{table*}

\section{Mass combinations}

\begin{figure*}[hbtp]
  \begin{center}
    \begin{tabular}{cc}
      \includegraphics[width=0.4\textwidth]{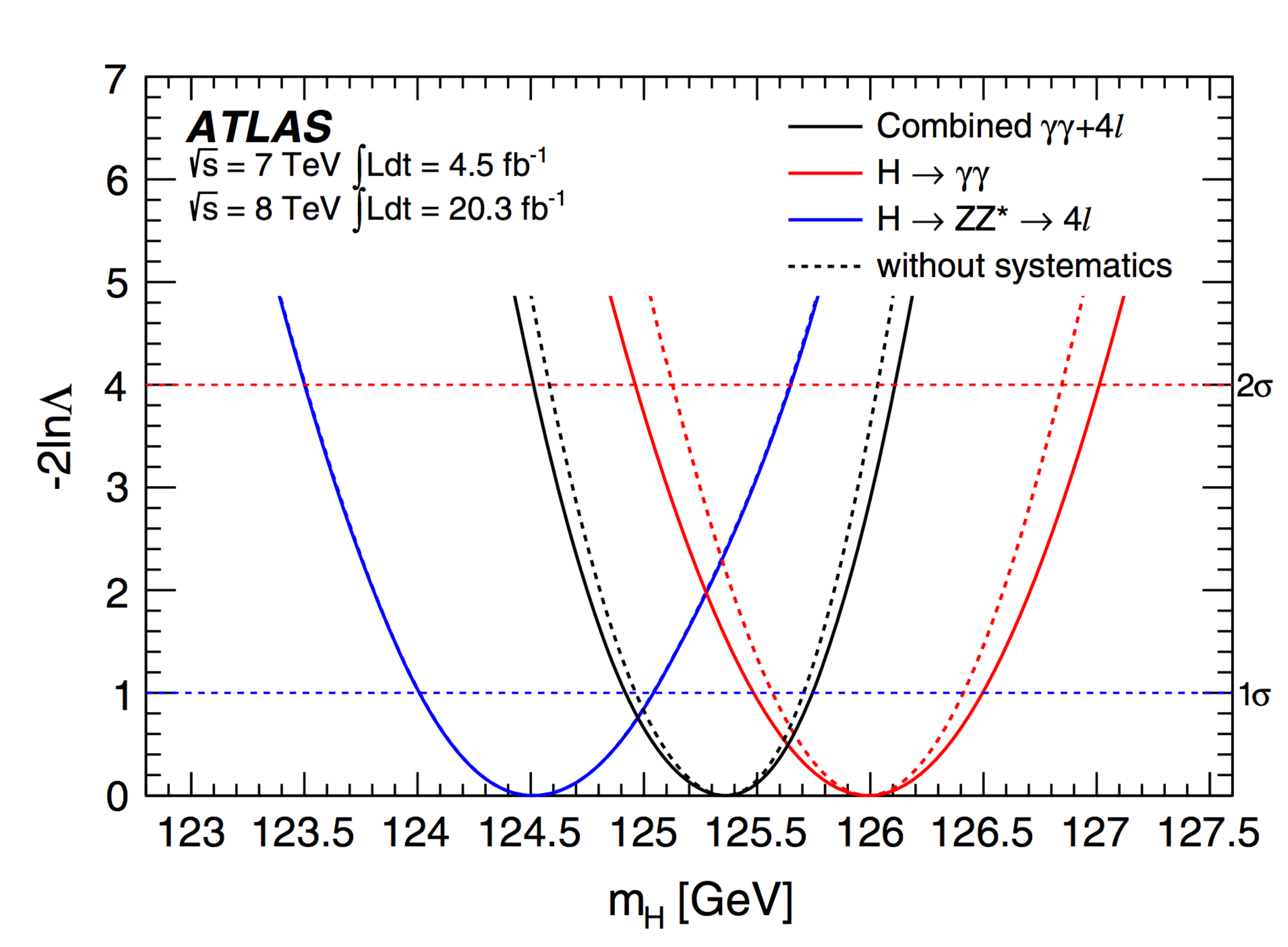} &
      \includegraphics[width=0.3\textwidth]{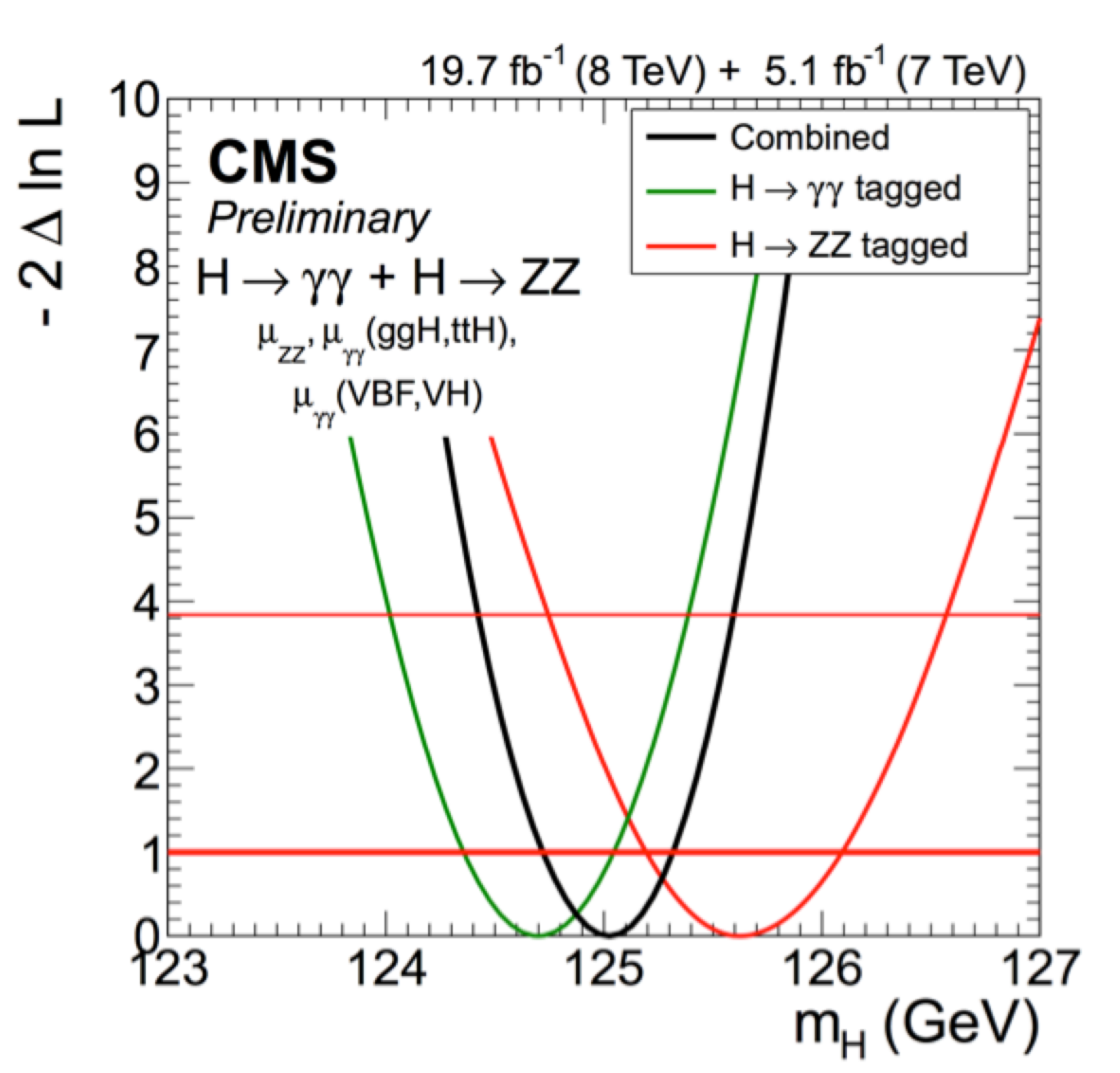}
    \end{tabular}
    \caption{The likelihood scans for the mass measurement combinations of the scalar boson from ATLAS \cite{ATLASCombo} (left) and CMS \cite{CMSCombo} (right) for the $\HToZZ$ and $\HToGG$ final states.}
    \label{fig:combination}
  \end{center}
\end{figure*}

Combining the mass measurements from the $\HToZZ$ and $\HToGG$ final states gives a good cross check of these measurements, which are within $1.2\sigma$ of each other.  Correlations between the electron and photon energy scales are taken into account when the combinations are performed.  The combinations for ATLAS and CMS are shown in table \ref{tab:combination}, and the likelihood scans are shown in figure \ref{fig:combination}.  The statistical uncertainties are larger than the systematic uncertainties.

\begin{table*}
  \begin{center}
    \begin{tabular}{ccccc}
      \hline
                                             & ATLAS                                           & CMS        \\
      \hline
      $\HToZZ$    & $124.51 \pm 0.52$ (stat) $\pm 0.06$ (syst) & $125.59 \pm {}^{+0.43}_{-0.41} $ (stat) $\pm {}^{+0.16}_{-0.18} $ (syst) \\
      $\HToGG$    & $125.98 \pm 0.42$ (stat) $\pm 0.28$ (syst) & $124.70 \pm 0.31$ (stat) $\pm 0.15$ (syst) \\
      \hline
      Combination & $125.36 \pm 0.37$ (stat) $\pm 0.18$ (syst) & $125.03 \pm 0.27$ (stat) $\pm 0.15$ (syst) \\
      \hline
    \end{tabular}
    \caption{The mass measurement combinations, in $\gev$, of the scalar boson from ATLAS \cite{ATLASCombo} and CMS \cite{CMSCombo} for the $\HToZZ$ and $\HToGG$ final states.  (stat) refers to statistical uncertainties and (syst) refers to systematic uncertainties.}
    \label{tab:combination}
  \end{center}
\end{table*}

\section{Future prospects}

Future prospects for the LHC and various linear colliders were presented at the Snowmass 2013 meeting \cite{Snowmass}.  Benchmarks projections are given for $\intL = 300 ~\fb^{-1}$, $3000 ~\fb^{-1}$.  Mass measurement uncertainties at the LHC will be dominated by lepton and photon energy and momentum scale uncertainties, which will decrease as statistics increase.  Optimistic projections of mass resolution at $\sqrt{s} = 14 ~\tev$ are $70 ~\mev$ for $\intL = 300 ~\fb^{-1}$, and $25 ~\mev$ for   $\intL = 3000 ~\fb^{-1}$.  (These values were presented at Snowmass 2013 and are not official ATLAS or CMS projections.)

In the $\HToGG$ final state there is a non-negligible shift in the mass peak, due to interference between standard model $gg\to\gamma\gamma$ and $gg\to\HToGG$.  This shift is of order $70 ~\mev$, depending upon the scalar boson width, and will become more significant as mass resolution in the $\HToGG$ final state improves.

\begin{figure*}[!hbtp]
  \begin{center}
    \includegraphics[width=0.75\textwidth]{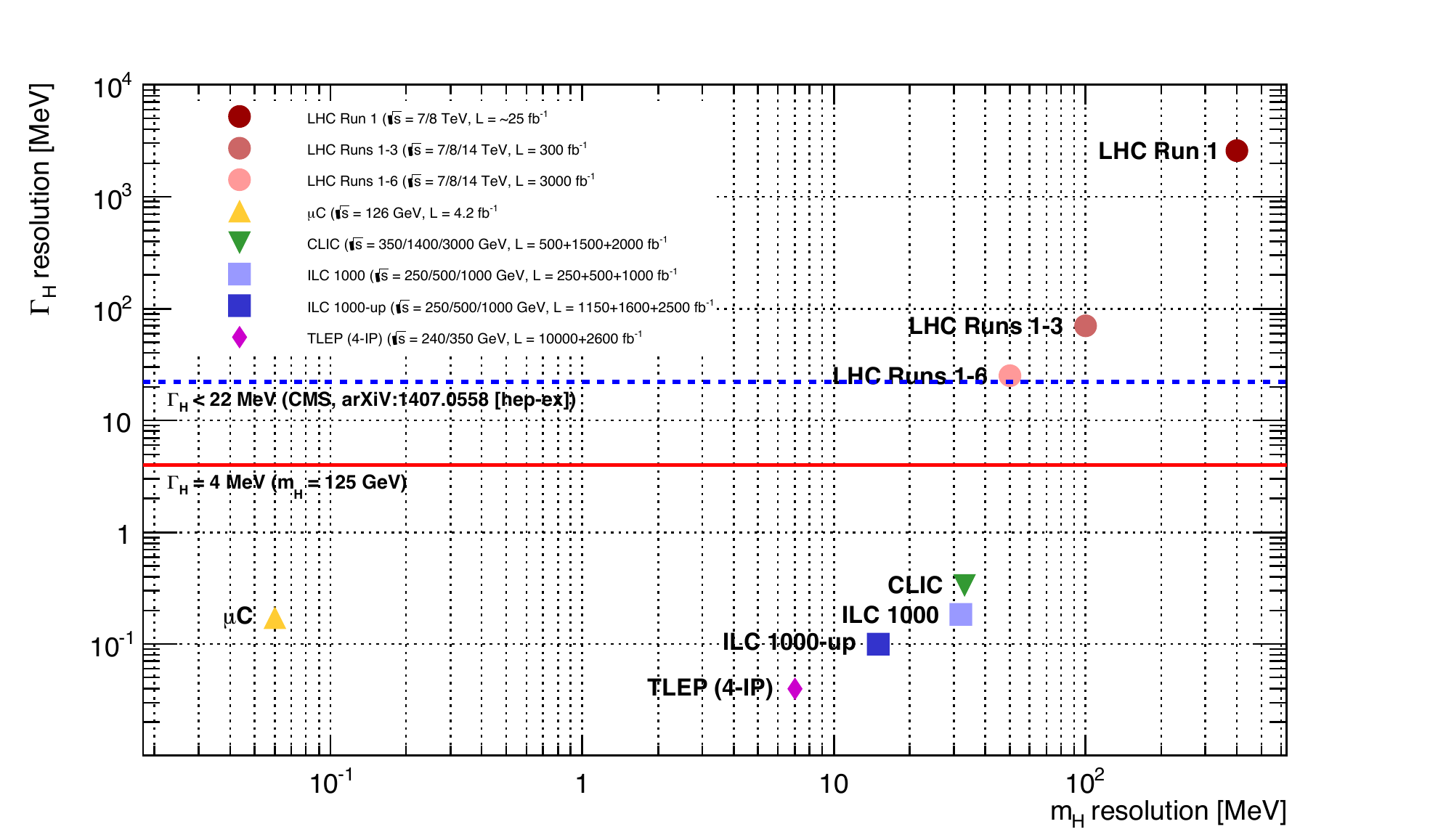}
    \caption{Future prospects for the anticipated mass and width resolution measurements of the scalar boson for various hadron and lepton colliders, as they were presented at the Snowmass 2013 meeting.}
    \label{fig:future}
  \end{center}
\end{figure*}

For $e^+e^-$ colliders the scalar boson will be produced mainly via the process $e^+e^-\to Z^{\star}\to ZH$, which was the final state searched for at the LEP collider.  For the $\mu^+\mu^-$ colliders the lineshape of the scalar boson can be measured directly as it was for the $Z$ boson at the LEP experiments.  Future prospects for the linear colliders are shown in table \ref{tab:future}, and the prospects of the mass resolution and width are shown in figure \ref{fig:future}.

\begin{table*}
  \begin{center}
    \begin{tabular}{ccccccc}
      \hline
       Facility                  & ILC500    & ILC1000   & ILC1000-up  & CLIC       & TLEP (4IP) & $\mu$C  \\
      \hline
      $\sqrt{s} (\gev)$          & $250/500$ & $250/500$ & $250/500$   & $350/1400$ & $240/350$  & $126$   \\
                                 &           & $/1000$   & $/1000$     & $/3000$    &            &         \\
      \hline
      $\intL (\fb^{-1})$         & $250+500$ & $250+500$ & $1150+1600$ & $500+1500$ & $10000$    & $4.2$   \\
                                 &           & $+1000$   & $+2500$     & $+2000$    & $+2600$    & $4.2$   \\
      \hline
      $m_H$ uncertainty $(\mev)$ & $32$      & $32$      & $15$        & $33$       & $7$        & $0.06$  \\
      \hline
      $\Gamma_H$ uncertainty     & $5.0\%$   & $4.6\%$   & $2.5\%$     & $8.4\%$    & $1.0\%$    & $4.3\%$ \\
      \hline
    \end{tabular}
    \caption{Future prospects for the anticipated mass and width resolution measurements of the scalar boson for various hadron and lepton colliders, as they were presented at the Snowmass 2013 meeting. \cite{Snowmass}.}
    \label{tab:future}
  \end{center}
\end{table*}

\section{Conclusion}

The ATLAS and CMS collaboration have measured the mass of the scalar boson with a resolution of less than $1\%$.  In both cases the conditions of the detector have been very challenging and studies have pushed the mass resolutions to the limit of what is currently achievable, limited by statistics.  Systematic uncertainties on the scalar boson mass are limited by the lepton and photon energy and momentum scale factors.  It is expected that the resolution will improve to $\sim 25 ~\mev$ with an integrated luminosity of $\intL = 3000 ~\fb^{-1}$ at the LHC.  With $\sim 10^7$ events per resonance the expected momentum scale factor uncertainties are of the order of $\sim 0.5\%$ per object, or better.  Further improvements in mass resolution will require a high energy lepton collider.

\end{document}